\documentstyle [12pt]{article}
\let\tilde=\widetilde

\def\one#1{#1^{\raise5pt\hbox{$\scriptstyle\!\!\!\!1$}}\,{}}
\def\two#1{#1^{\raise5pt\hbox{$\scriptstyle\!\!\!\!2$}}\,{}}
\def\three#1{#1^{\raise5pt\hbox{$\scriptstyle\!\!\!\!3$}}\,{}}
\def\phi{\varphi}

\def\a{\alpha}

\def\D{\Delta}

\def\s{\sigma}

\def\comment#1{}

\def\id{\hbox{{1}\kern-.25em\hbox{\rm l}}}
\def\beq{\begin{equation}}
\def\eeq{\end{equation}}
\def\be{\begin{displaymath}}
\def\ee{\end{displaymath}}
\def\bmat{\left(\begin{array}}
\def\emat{\end{array}\right)}
\def\bds{\begin{description}}
\def\eds{\end{description}}
\def\?{(?)\marginpar{|?}}

\def\half{\frac{1}{2}}

\newtheorem{theo}{Theorem}
\newtheorem{prop}{Proposition}

\def\dd{\partial}
\def\H{{\cal H}}
\def\T{{\cal T}}
\def\D{{\cal D}}
\def\endproof{\hfill\rule{2mm}{2mm}}

\newfont{\bbd}{msbm10 scaled\magstep1} 
\def\R{\hbox{\bbd R}}                  
\def\Z{\hbox{\bbd Z}}
\newfont{\frak}{eufm10 scaled\magstep1}
\begin{document}
\begin{flushright}
\sf UTMS 95-10 \\
    UAMS 95-06 \\
    solv-int/9508002
\end{flushright}
\begin{center}{\Large\bf
Separation of variables in the \\
$A_2$ type Jack polynomials }
\vspace{.3cm}

V.B.~Kuznetsov\vspace{.2cm}\\
Faculteit voor Wiskunde en Informatica, Universiteit van Amsterdam,\\
Plantage Muidegracht 24, 1018 TV Amsterdam, The Netherlands${}^1$
\vspace{.3cm}\\
E.K.~Sklyanin\vspace{.2cm}\\
Department of Mathematical Sciences, The University of Tokyo,\\
7-3-1 Hongo, Bunkyo-ku, Tokyo 113, Japan${}^2$
\end{center}
\footnotetext[1]{On leave from Department of Mathematical
and Computational Physics, Institute of Physics, St.~Petersburg
University, St.~Petersburg 198904, Russia.}
\footnotetext[2]{On leave from Steklov Mathematical Institute,
 Fontanka 27, St.Petersburg 191011, Russia.}
\vspace{.8cm}
\centerline{\bf Abstract}
An integral operator $M$ is constructed performing a separation
of variables for the 3-particle quantum Calogero-Sutherland (CS) model.
Under the action of $M$ the CS eigenfunctions (Jack polynomials
for the root system $A_2$) are transformed to the factorized form
$\phi(y_1)\phi(y_2)$, where $\phi(y)$ is a trigonometric polynomial
of one variable
expressed in terms of the ${}_3F_2$ hypergeometric series. The
inversion of $M$ produces a new integral representation for the
$A_2$ Jack polynomials.
\vskip 1cm
\section{Quantum Calogero-Sutherland model}\label{QCSM}

Define $N$ differential operators $\{H_k\}_{k=1}^N$,
acting on functions of $N$ variables $\vec q=\{q_1,\ldots,q_N\}$
and depending on a parameter $g$, by the formula \cite{Oshima-Seki}
\beq\label{eq:def-Hk}
  H_k=\sum_{0\leq l\leq\frac{k}{2}}\sum_{\s\in\mbox{\frak S}_N}
   \frac{1}{\# G(l,k-2l)}D^\s_{l,k-2l}
\eeq
where
\beq
 D_{m,n}=u(q_1-q_2)u(q_3-q_4)\ldots u(q_{2m-1}-q_{2m})
\frac{(-i)^n\dd^n}{\dd q_{2m+1}\dd q_{2m+2}\ldots\dd q_{2m+n}}.
\eeq

Here we denote $u(q)=-g(g-1)/\sin^2 q$, whereas $\mbox{\frak S}_N$ is the
permutation
group of the set $\{1,\ldots,N\}$, and $G(m,n)=\{\s\in\mbox{\frak S}_N |
D^\s_{m,n}=D_{m,n}\}$.

Note that, when $g\rightarrow0$, the operators $H_k$ behave as
\beq
 H_k=(-i)^k\sum_{j_1<\cdots<j_k}\frac{\dd^k}{\dd q_{j_1}\ldots\dd q_{j_k}}
+{\cal O}(g),
\eeq
providing thus a one-parameter deformation of the elementary symmetric
polynomials in $\dd/\dd q_j$.

It is known  \cite{Oshima-Seki} that the operators $H_k$ generate
a commutative ring which contains, in particular, the quantum
Calogero-Sutherland \cite{Calog,Suth,OP1,OP2} Hamiltonian
\beq
 H=\half H_1^2-H_2=-\half\sum_{j=1}^N\frac{\dd^2}{\dd q_j^2}
   +\sum_{j_1<j_2}\frac{g(g-1)}{\sin^2(q_{j_1}-q_{j_2})}.
\eeq

To describe the quantum problem more
precisely, define the space of quantum states $\H^{(N)}$ as the
complex Hilbert space of functions $\Psi$ on the torus
$T^{(N)}=\R^N/\pi\Z^N\ni\vec q$ which are symmetric w.r.t.\ the permutations
of $q_j$, the scalar product being defined as
\beq\label{eq:norm}
 \left<\Psi,\Phi\right>=\int_0^\pi \!dq_1\ldots\int_0^\pi \!dq_N
\,\bar\Psi(\vec q)\Phi(\vec q).
\eeq

Note that for the real $g$
the operators (\ref{eq:def-Hk}) are formally Hermitian
w.r.t.\ the above sesquilinear form.
Let the vacuum (ground state) function $\Omega$ be defined as
\beq\label{eq:Omega}
 \Omega(\vec q)=\left|\prod_{j<k}\sin(q_j-q_k)\right|^g.
\eeq

Though $\Omega\in\H^{(N)}$ for $g>-\half$, we shall assume a more
strong condition $g>0$ which simplifies description of the eigenvectors.
Let $\T^{(N)}$ be the space
of symmetric trigonometric polynomials in variables $\vec q$,
that is the symmetric Laurent polynomials in variables
$t_j=e^{2iq_j}$. The simplest way to fix the ``boundary
conditions'' for the operators $H_k$ is to restrict them first on the
dense linear subset $\D_g^{(N)}=\Omega\T^{(N)}\subset\H^{(N)}$.
Since $\D_g^{(N)}$ consists
of common analytical vectors of operators $H_k$, the latter can be
extended uniquely to commuting self-adjoint operators in $\H^{(N)}$.

The complete set of orthogonal eigenvectors to the self-adjoint $H_k$
\beq
 H_k\Psi_{\vec n}=h_k\Psi_{\vec n}
\eeq
is well known \cite{Suth,OP2}.
The eigenvectors are parametrized by the sequences
$\vec n=\{n_1\leq n_2\leq\ldots \leq n_N\}$
of integers $n_j\in\Z$. The corresponding eigenvalues $h_k$ are
\beq\label{eq:def-hm}
 h_k=2^k\sum_{j_1<\cdots<j_k}m_{j_1}\ldots m_{j_k},
\qquad  m_j=n_j+g\left(j-\frac{N+1}{2}\right).
 \eeq

The eigenfunctions allow the factorization
\beq\label{eq:Psi-Om}
 \Psi_{\vec n}({\vec q})=\Omega(\vec q)J_{\vec n}({\vec q}),
\qquad J_{\vec n}\in\T^{(N)}.
\eeq

In particular, for the ground state $\Omega=\Psi_{0\ldots0}$
and $J_{0\ldots0}=1$.
The symmetric trigonometric polynomials $J_{\vec n}$ are known
as Jack polynomials corresponding to the root system $A_{N-1}$
or simple Lie algebra $sl_N$, see \cite{Macd} and also \cite{Koor74} for
the $A_2$ case.
Our notation differs slightly from the conventional one: our
parameter $g$ relates to $\a$ used in \cite{Macd}
as $g=\a^{-1}$,
and we do not impose the restriction $n_j\geq 0$.

The problem of finding square integrable eigenfunctions
$\Psi\in\H^{(N)}$
of the operators $H_k$ turns out thus to be equivalent to
the purely algebraic problem of finding the polynomial eigenfunctions
$J\in\T^{(N)}$ of the differential operators $\tilde H_k$
obtained by conjugation of $H_k$ with $\Omega$
\beq\label{eq:def-tHk}
 \tilde H_k=\Omega^{-1}H_k\Omega.
\eeq

Jack polynomials can be considered as a one-parametric deformation
of elementary symmetric polynomials
$ S_{\vec n}(\vec q)=\sum t_1^{\nu_1}\ldots t_N^{\nu_N} $
where the sum is taken over all distinct permutations $\vec \nu$
of $\vec n$, such that
\beq\label{eq:norm-J}
 J_{\vec n}=S_{\vec n}+\sum_{\vec n'\preceq \vec n}
\kappa_{\vec n\vec n'}S_{\vec n'},
\eeq
where $\kappa_{\vec n\vec n'}$ is a rational function in $g$ vanishing
for $g=0$, and the dominant order for sequences $\vec n$ is defined as
\beq\label{eq:dominant}
 \vec n\succeq\vec n'  \quad\Longleftrightarrow\quad
\left\{\sum_{j=1}^N n_j=\sum_{j=1}^N n_j^\prime; \quad
  \sum_{j=k}^N n_j \geq \sum_{j=k}^N n_j^\prime, \quad
  k=2,\ldots,N \right\}
\eeq

Another important property of Jack polynomials is the orthogonality
with the weight $\Omega^2$,
\beq
\int_0^\pi \!dq_1\ldots\int_0^\pi \!dq_N \,
\bar J_{\vec n}(\vec q)J_{\vec n'}(\vec q)\Omega^2(\vec q)=0,
\quad \vec n\neq \vec n'
 \eeq

For the generalization of Jack polynomials for other root systems see
\cite{HeckOp}.

\section{Separation of variables: conjectures}\label{SoV}

In the classical case, when the differentiation $-i\dd/\dd q_j$ is
replaced by the momentum $p_j$ canonically conjugated to $q_j$,
the Calogero-Sutherland system is completely integrable in the
Liouville's sense  \cite{Calog,OP1}. It is thus natural
to speak of its quantum version described above as a quantum integrable
system. The common property to be expected from an integrable system
(classical or quantum one) is the {\it separability of variables}
\cite{Kaln,Kuz,Skl:32,Skl:38} which suggests the following
conjecture.

{\bf Conjecture 1.} {\it There exists a linear operator
\beq
 K:\Psi_{\vec n}({\vec q})\longmapsto
   \tilde\Psi_{\vec n}(y_1,\ldots,y_{N-1};Q)
\eeq
such that any eigenfunction $\Psi_{\vec n}$ is transformed into
the factorized function
\beq
\label{eq:factor-Psi}
 \tilde\Psi_{\vec n}(y_1,\ldots,y_{N-1};Q)=
 e^{ih_1Q}\prod_{k=1}^{N-1}\psi_{\vec n}(y_k).
\eeq
}

The distinguished variable $Q\equiv q_N$ is simply the coordinate canonically
conjugated to the total momentum $H_1$.

The study of the low-dimensional cases $N=2,3$
allows to formulate an even more detailed conjecture about
the structure of the separated eigenfunction $\tilde\Psi$.

{\bf Conjecture 2.} {\it The factor $\psi_{\vec n}(y)$ in
(\ref{eq:factor-Psi}) allows further factorization
\beq\label{eq:fact-psi}
 \psi_{\vec n}(y)=(\sin y)^{(N-1)g}\phi_{\vec n}(y)
\eeq
where $\phi_{\vec n}(y)$ is a Laurent polynomial in $t=e^{2iy}$
\beq\label{eq:phi-Laurent}
 \phi_{\vec n}(y)=\sum_{k=n_1}^{n_N}t^k c_k(\vec n;g).
\eeq

The coefficients $c_k(\vec n;g)$ are rational functions of
$k$, $n_j$ and $g$. Moreover, $\phi_{\vec n}(y)$
can be expressed
explicitely in terms of the hypergeometric function ${}_N F_{N-1}$
as
\beq\label{eq:phi-hgf}
 \phi_{\vec n}(y)=t^{n_1}(1-t)^{1-Ng}
{}_N F_{N-1}(a_1,\ldots,a_N;b_1,\ldots,b_{N-1};t)
\eeq
where
\beq\label{eq:def-ab}
 a_j=n_1-n_{N-j+1}+1-(N-j+1)g, \qquad
 b_j=a_j+g,
\eeq
\beq\label{eq:def-hgf}
 {}_N F_{N-1}(a_1,\ldots,a_N;b_1,\ldots,b_{N-1};t)=
\sum_{k=0}^\infty\frac{(a_1)_k\ldots(a_N)_k t^k}%
{(b_1)_k\ldots(b_{N-1})_k k!},
\eeq
and $(a)_k$ is the standard Pochhammer symbol:
\beq
 (a)_0=1, \qquad (a)_k=a(a+1)\ldots(a+k-1)=
     \frac{\Gamma(a+k)}{\Gamma(a)}.
\eeq
}

The conjectures 1 and 2 are proved
in the next section for the $N=2$ case
and in the sections \ref{int-tr} and \ref{sep-eq} for the $N=3$ case.
Section \ref{sep-eq} contains also
a more detailed discussion of the conjecture 2 for $N>3$, see theorem
\ref{Vadim}.
Further support to the conjectures is given by the study
of the case $g=1$ when Jack polynomials degenerate into Schur functions
(section \ref{Schur}).

\section{$A_1$ case}

It is a well known fact that in the $A_1$ case Jack polynomials
are reduced to hypergeometric polynomials of one variable
\cite{HeckOp}.
Nevertheless, we review the derivation briefly in order to
prepare the stage for the discussion of the $A_2$ case.

For $N=2$ the commuting operators (\ref{eq:def-Hk}) are
\beq
  H_1=-i(\dd_1+\dd_2), \qquad
  H_2=-\dd_1\dd_2-g(g-1)\sin^{-2}q_{12}.
\eeq
(we denote $\dd_j=\dd/\dd q_j$ and $q_{jk}=q_j-q_k$).
Respectively,
\begin{eqnarray*}
  \tilde H_1=-i(\dd_1+\dd_2), \qquad
  \tilde H_2=-\dd_1\dd_2+g\cot q_{12}(\dd_1-\dd_2)-g^2,
\end{eqnarray*}
the vacuum vector  being
\beq
\Omega(\vec q)=\left|\sin q_{12}\right|^g.
\eeq

The eigenvectors $\Psi_{\vec n}$, resp.\ $J_{\vec n}$,
according to (\ref{eq:def-hm}), are parametrized
by the pairs of integers $\vec n=\{n_1\leq n_2\}$, the corresponding
eigenvalues being
\beq
   h_1=2(m_1+m_2)=2(n_1+n_2), \qquad
   h_2=4m_1m_2=(2n_1-g)(2n_2+g)
\eeq
where
\beq
 m_1=n_1-\frac{g}{2}, \qquad m_2=n_2+\frac{g}{2}.
\eeq

The separation of variables is given by the simple change of coordinates
\beq
 K:\Psi(q_1,q_2) \longmapsto \tilde\Psi(y,Q)=\Psi(y+Q,Q).
\eeq

Actually, the calculations would be simpler for the more
symmetric definition $Q=(q_1+q_2)/2$  rather than $Q=q_2$ but
we wish to preserve here the coherence of notation for
the study of $N=3$ case.

The spectral problem $H_k\Psi=h_k\Psi$ rewritten in terms
of the function $\tilde\Psi$ reads
\beq
  \left[\dd_Q-ih_1\right]\tilde\Psi=0, \qquad
  \left[\dd^2_y-\dd_y\dd_Q-\frac{g(g-1)}{\sin^2 y}-h_2\right]\tilde\Psi=0,
\eeq
allowing immediate separation of variables of the form (\ref{eq:factor-Psi})
\beq
 \tilde\Psi(y,Q)=e^{ih_1Q}\psi(y),
\eeq
the function $\psi$ satisfying the second order differential
equation
\beq
\left[\dd_y^2-ih_1\dd_y-\left(h_2+\frac{g(g-1)}{\sin^2y}\right)\right]\psi
=0
\eeq
which, via the transformation $\psi(y)=\sin^g y\,\phi(y)$,
can be rewritten  as
\beq
\left[\dd_y^2+(2g\cot y-ih_1)\dd_y-(g^2+igh_1\cot y+h_2)\right]\phi=0.
\eeq

The last equation, after the substitution $t=e^{2iy}$, is reduced
to the standard Fuchsian form
\beq\label{eq:Fuchs2}
\left[\dd_t^2+\left(
 -\frac{g-1+\half h_1}{t}+\frac{2g}{t-1}\right)\dd_t
+\left(
 \frac{\frac{1}{4}(g^2+gh_1+h_2)}{t^2}-\frac{\half gh_1}{t(t-1)}
\right)\right]\phi=0.
\eeq

The equation (\ref{eq:Fuchs2}) has 3 regular singularities: $\{0,1,\infty\}$
with the characteristic exponents:
$$ \begin{array}{lll}
 t\sim 1 & \phi\sim(t-1)^\mu & \mu\in\{-2g+1,0\} \\
 t\sim 0 & \phi\sim t^\rho & \rho\in\{n_1,n_2+g\} \\
 t\sim\infty & \phi\sim t^{-\s} & -\s\in\{n_1-g,n_2\}
\end{array}
$$

Moreover, by the substitution $\phi(t)\!=\!t^{n_1}(1-t)^{1-2g}f(t)$
the equation (\ref{eq:Fuchs2}) is reduced to the standard
hypergeometric equation
\beq
 \left[t\dd_t(t\dd_t+b_1-1)-t(t\dd_t+a_1)(t\dd_t+a_2)\right]f=0,
\eeq
the parameters $a_1$, $a_2$, $b_1$ being given by the formulas
(\ref{eq:def-ab}) which for $N=2$ read
\beq\label{eq:def-ab2}
 a_1=n_1-n_2+1-2g, \qquad
 a_2=1-g, \qquad
 b_1=n_1-n_2+1-g.
\eeq

{}From $J_{n_1n_2}\in\T^{(2)}$ it follows immediately that the
corresponding $\phi_{n_1n_2}(t)$ is a Laurent polynomial in $t$.

\begin{prop}
 The Laurent polynomial $\phi_{n_1n_2}(t)$ is given, up to
a constant factor, by the formula (\ref{eq:phi-hgf}) which,
for $N=2$ takes the form
\beq\label{eq:phi-hgf2}
  \phi_{n_1n_2}(t)=t^{n_1}(1-t)^{1-2g}\,
{}_2F_1(a_1,a_2;b_1;t)
\eeq
the parameters $a_1$, $a_2$, $b_1$ being given by (\ref{eq:def-ab2}).
\end{prop}

{\bf Proof.}
Define the function $F_{n_1n_2}(t)$ by the right hand side of the
formula (\ref{eq:phi-hgf2}). Strictly speaking, the hypergeometric
series converges only for $|t|<1$ but in few moments we shall see
that $F_{n_1n_2}(t)$ continues analytically to the whole complex plane.
Using the well known formula
$$
  (1-t)^{a+b-c}\,{}_2F_1(a,b;c;t)={}_2F_1(c-a,c-b;c;t)
$$
we can rewrite $F_{n_1n_2}(t)$ as folllows
$$
F_{n_1n_2}(t)=t^{n_1}\,{}_2F_1(n_1-n_2,g;n_1-n_2+1-g;t)
$$

It is easy to observe now that the hypergeometric series
in the right hand side terminates for integer $\{n_1\leq n_2\}$
and $F_{n_1n_2}$ is thus a Laurent polynomial
$$
    F_{n_1n_2}=\sum_{k=n_1}^{n_2}t^k c_k(\vec n;g),
$$
of the form (\ref{eq:phi-Laurent}).
Since $F_{n_1n_2}$ satisfies the same differential equation
(\ref{eq:Fuchs2}) as
$\phi_{n_1n_2}$ and the linearly independent solution to
(\ref{eq:Fuchs2}) is obviously not polynomial, the functions
$F_{n_1n_2}(t)$ and $\phi_{n_1n_2}(t)$ are identical up to a
constant factor, which finishes the proof of the proposition
and of the conjectures 1 and 2 for $N=2$.\endproof

\section{$A_2$ case: Integral transform}\label{int-tr}

For $N=3$ the commuting differential operators (\ref{eq:def-Hk})
read
\begin{eqnarray*}
 H_1 & = & -i(\dd_1+\dd_2+\dd_3), \\
 H_2 & = & -(\dd_1\dd_2+\dd_1\dd_3+\dd_2\dd_3)
      -g(g-1)\left(\sin^{-2}q_{12}+\sin^{-2}q_{13}+\sin^{-2}q_{23}
\right), \\
 H_3 & = & i\dd_1\dd_2\dd_3
       +ig(g-1)\left(\sin^{-2}q_{23}\,\dd_1
                    +\sin^{-2}q_{13}\,\dd_2+\sin^{-2}q_{12}\,\dd_3
\right), \\
\end{eqnarray*}
and, respectively,
\begin{eqnarray*}
 \tilde H_1&=&-i(\dd_1+\dd_2+\dd_3) \\
 \tilde H_2&=&-(\dd_1\dd_2+\dd_1\dd_3+\dd_2\dd_3) \\
         &&g{[}\cot q_{12}(\dd_1-\dd_2)+\cot q_{13}(\dd_1-\dd_3)
               +\cot q_{23}(\dd_2-\dd_3){]} \\
         &&-4g^2 \\
 \tilde H_3&=& i\dd_1\dd_2\dd_3 \\
     &&-ig{[}\cot q_{12}(\dd_1-\dd_2)\dd_3
             +\cot q_{13}(\dd_1-\dd_3)\dd_2
             +\cot q_{23}(\dd_2-\dd_3)\dd_1 {]}\\
     &&+2ig^2{[} (1+\cot q_{12}\cot q_{13})\dd_1
                +(1-\cot q_{12}\cot q_{23})\dd_2
                +(1+\cot q_{13}\cot q_{23})\dd_3 {]}
\end{eqnarray*}
the vacuum function being
\beq
\Omega(\vec q)=\left|\sin q_{12}\sin q_{13}\sin q_{23}\right|^g.
\eeq

The eigenvectors $\Psi_{\vec n}$, resp.\ $J_{\vec n}$,
according to (\ref{eq:def-hm}), are parametrized
by the triplets of integers $\{n_1\leq n_2\leq n_3\}\in\Z^3$,
the corresponding eigenvalues being
\beq\label{eq:def-h3}
 h_1=2(m_1+m_2+m_3), \quad
 h_2=4(m_1m_2+m_1m_3+m_2m_3), \quad
 h_3=8m_1m_2m_3,
\eeq
where,
\beq\label{eq:def-m3}
      m_1=n_1-g, \qquad
      m_2=n_2, \qquad
      m_3=n_3+g.
\eeq

The structure of the operator $K$ performing separation
of variables in the $A_2$ case is more complicated than
in the $A_1$ case. In contrast with the $A_1$ case, $K$ is
given by an integral operator rather then by simple change of
coordinates. To describe $K$, let us introduce the following notation.
$$  x_1=q_1-q_3, \qquad x_2=q_2-q_3, \qquad Q=q_3, $$
$$ x_\pm=x_1\pm x_2, \qquad y_\pm=y_1\pm y_2. $$

We shall study the action of $K$ locally, assuming
that $q_1>q_2>q_3$ and hence $x_+>x_-$.

The operator $K:\Psi(q_1,q_2,q_3)\mapsto\tilde\Psi(y_1,y_2;Q)$ is defined
as an integral operator
\beq\label{eq:def-K}
  \tilde\Psi(y_1,y_2;Q)=\int_{y_-}^{y_+}d\xi\,{\cal K}(y_1,y_2;\xi)
  \Psi\left(\frac{y_++\xi}{2}+Q,\frac{y_+-\xi}{2}+Q,Q\right)
\eeq
with the kernel
\beq\label{eq:kernel-K}
 {\cal K}=\kappa\left[\frac{
   \sin\left(\frac{\displaystyle \xi+y_-}{\displaystyle 2}\right)
   \sin\left(\frac{\displaystyle \xi-y_-}{\displaystyle 2}\right)
   \sin\left(\frac{\displaystyle y_++\xi}{\displaystyle 2}\right)
   \sin\left(\frac{\displaystyle y_+-\xi}{\displaystyle 2}\right)}{
 \sin y_1\sin y_2\sin \xi}\right]^{\displaystyle g-1}
\eeq
where $\kappa$ is a normalization coefficient to be fixed later.
It is assumed in (\ref{eq:def-K}) and (\ref{eq:kernel-K})
that $y_-<x_-=\xi<y_+=x_+$.
The integral converges when $g>0$ which will always be assumed
henceforth.

The motivation for such a choice of $K$ takes its origin
from considering the problem in the classical limit
$(g\rightarrow\infty)$ where there exists effective prescription
for constructing a separation of variables for an integrable system
from the poles of the so-called Baker-Akhiezer function.
See \cite{Skl:38}, \S7,  for a detailed explanation.

\begin{theo}\label{Psi-diff-eq}
 Let $H_k\Psi_{n_1n_2n_3}=h_k\Psi_{n_1n_2n_3}$.
Then the function $\tilde\Psi_{\vec n}=K\Psi_{\vec n}$
satisfies the differential equations
\beq
 {\cal Q}\tilde\Psi_{\vec n}=0,\qquad
 {\cal Y}_j\tilde\Psi_{\vec n}=0, \quad j=1,2
\eeq
where
\beq
 {\cal Q}=-i\dd_Q-h_1,
\eeq
\begin{eqnarray}
\lefteqn{
 {\cal Y}_j=i\dd_{y_j}^3+h_1\dd_{y_j}^2
-i\left(h_2+3\frac{g(g-1)}{\sin^2{y_j}}\right)\dd_{y_j}} \nonumber \\
&& -\left(h_3+\frac{g(g-1)}{\sin^2{y_j}}h_1
    +2ig(g-1)(g-2)\frac{\cos {y_j}}{\sin^3 {y_j}}\right).
\end{eqnarray}
\end{theo}

The proof is based on the following proposition.
\begin{prop}\label{K:q-char-eq}
The kernel $K$ satisfies the differential equations
$$ [-i\dd_Q-H_1^*]K=0, $$
\begin{eqnarray*}
\lefteqn{ \left[i\dd^3_{y_j}+H_1^*\dd^2_{y_j}
   -i\left(H_2^*+\frac{3g(g-1)}{\sin^2 y_j}\right)\dd_{y_j} \right.} \\
&& \left.   -\left(H_3^*+H_1^*\frac{g(g-1)}{\sin^2 y_j}
  +2ig(g-1)(g-2)\frac{\cos y_j}{\sin^3 y_j}\right)\right]K=0,
\end{eqnarray*}
where $H_n^*$ is the Lagrange adjoint of $H_n$
$$ \int\phi(q)(H\psi)(q)\,dq=\int(H^*\phi)(q)\psi(q)\,dq $$
\begin{eqnarray*}
 H^*_1&=&i(\dd_{q_1}+\dd_{q_2}+\dd_{q_3}), \\
 H^*_2&=&-\dd_{q_1}\dd_{q_2}-\dd_{q_1}\dd_{q_3}-\dd_{q_2}\dd_{q_3}
   -g(g-1){[}\sin^{-2}q_{12}+\sin^{-2}q_{13}+\sin^{-2}q_{23}{]}, \\
 H^*_3&=&-i\dd_{q_1}\dd_{q_2}\dd_{q_3}-ig(g-1)
    {[}\sin^{-2}q_{23}\,\dd_{q_1}+\sin^{-2}q_{13}\,\dd_{q_2}
     +\sin^{-2}q_{12}\dd_{q_3}{]}.
\end{eqnarray*}
\end{prop}

The proof is given by a direct, though tedious calculation.

To complete the proof of the theorem \ref{Psi-diff-eq}, consider
the expressions ${\cal Q}K\Psi_{\vec n}$ and ${\cal Y}_jK\Psi_{\vec n}$
using the formulas (\ref{eq:def-K}) and (\ref{eq:kernel-K}) for $K$.
The idea is to use
the fact that $\Psi_{\vec n}$ is an eigenfunction of
$H_k$ and replace $h_k\Psi_{\vec n}$ by $H_k\Psi_{\vec n}$.
After integration by parts in the variable $\xi$ the operators
$H_k$ are replaced by their adjoints $H_k^*$ and the result is zero
by virtue of proposition \ref{K:q-char-eq}.

The caution is needed however when handling the limits of integration
$y_\pm$ in (\ref{eq:def-K}). The following argument allows to
circumvent the problem of boundary terms. One can hide the limits
of integration into the definition of the kernel ${\cal K}$
considering the factors containing $(\xi-y_\pm)$ as the generalized
functions similar to $x_+^\lambda$, see \cite{G-Sh}. It is known
that $x_+^\lambda$ defined through the linear functional
$$ \left<f,x_+^\lambda\right>=\int_0^\infty dx\, f(x)x_+^\lambda
$$
is analytic in $\lambda$ on the complex plane excluding
the poles $x=-1,-2,\ldots$
and can be differentiated just as usual power function
$\dd_xx_+^\lambda=\lambda x_+^{\lambda-1}$. Therefore, we can safely
ignore the boundary of integral (\ref{eq:def-K}) while
integrating by parts. The only possible obstacle may
present the integer points $g=1,2,3$ (no more, since we need to
differentiate ${\cal K}$ maximum 3 times) where the boundary
may contribute delta-function terms. The direct calculation shows,
however, that all such terms cancel.   \endproof

The following theorem validates the conjectures 1 and 2 for the
$A_2$ case.

\begin{theo}\label{Psi3-fact}
The function $\tilde\Psi_{n_1n_2n_3}$ is factorized
\beq
 \tilde\Psi_{n_1n_2n_3}(y_1,y_2;Q)=
e^{ih_1Q}\psi_{n_1n_2n_3}(y_1)\psi_{n_1n_2n_3}(y_2)
\eeq
according to (\ref{eq:factor-Psi}).
The separated function $\psi_{n_1n_2n_3}(y_2)$
has the structure (\ref{eq:fact-psi}).
\end{theo}

Note that, by virtue of the theorem \ref{Psi-diff-eq}, the function
$\tilde\Psi_{\vec n}(y_1,y_2;Q)$ satisfies an ordinary differential
equation in each variable. Since
${\cal Q}f=0$ is a first order differential equation
having a unique, up to a constant factor, solution $f(Q)=e^{ih_1Q}$,
the dependence on $Q$ is factorized. However, the differential
equations ${\cal Y}_j\psi(y_j)=0$ are of third order and have
three linearly independent solutions. To prove the theorem
\ref{Psi3-fact} one needs thus to study the ordinary
differential equation
\begin{eqnarray}
\lefteqn{
\left[ i\dd_y^3+h_1\dd_y^2
-i\left(h_2+3\frac{g(g-1)}{\sin^2y}\right)\dd_y\right.} \nonumber \\
&&  -\left.\left(h_3+\frac{g(g-1)}{\sin^2y}h_1
    +2ig(g-1)(g-2)\frac{\cos y}{\sin^3 y}\right)\right]\psi=0.
\label{eq:sep-eq3}
\end{eqnarray}
and to select its special solution corresponding to $\tilde\Psi$.

The proof will take several steps. First, let us eliminate from $\Psi$ and
$\tilde\Psi$ the vacuum factors $\Omega$, see (\ref{eq:Psi-Om}),
and, respectively
\beq\label{eq:tPsi-om}
   \tilde\Psi(y_1,y_2;Q)=\omega(y_1)\omega(y_2)\tilde J(y_1,y_2;Q),
 \qquad   \omega(y)=\sin^{2g}y.
\eeq

Conjugating the operator $K$ with the vacuum factors
\beq\label{eq:conj-K}
  M=\omega_1^{-1}\omega_2^{-1}K\Omega:J\mapsto\tilde J
\eeq
we obtain the integral operator
\beq\label{eq:def-M}
   \tilde J(y_1,y_2;Q)=\int_{y_-}^{y_+}d\xi\,{\cal M}(y_1,y_2;\xi)
  J\left(\frac{y_++\xi}{2}+Q,\frac{y_+-\xi}{2}+Q,Q\right)
\eeq
with the kernel
\begin{eqnarray}\label{eq:def-ker-M}
\lefteqn{
 {\cal M}(y_1,y_2;\xi)={\cal K}(y_1,y_2;\xi)
\frac{\Omega\left(\frac{y_++\xi}{2}+Q,\frac{y_+-\xi}{2}+Q,Q\right)}%
{\omega(y_1)\omega(y_2)} } \nonumber \\
&&= \kappa\sin \xi\frac{
 \left[
   \sin\left(\frac{\xi+y_-}{2}\right)
   \sin\left(\frac{\xi-y_-}{2}\right)
 \right]^{g-1}
 \left[
   \sin\left(\frac{y_++\xi}{2}\right)
   \sin\left(\frac{y_+-\xi}{2}\right)
 \right]^{2g-1}}{\left[
 \sin y_1\sin y_2\right]^{3g-1}}.
\end{eqnarray}

\begin{prop}\label{M:poly-poly}
Let $S$ be a trigonometric
polynomial in $q_j$, i.e.\ Laurent polynomial in $t_j=e^{2iq_j}$,
which is symmetric w.r.t.\ the transpositon $q_1\leftrightarrow q_2$.
 Then $\tilde S=MS$ is a trigonometric polynomial symmetric w.r.t.\
$y_1\leftrightarrow y_2$.
\end{prop}

{\bf Proof.} It is more convenient to use variables $x_\pm$, $Q$
and, respectively, $y_\pm$, $Q$. Since the kernel ${\cal M}$
does not depend on $Q$ it is safe to omit the dependence on $Q$ in $S$.
The polynomiality and symmetry of $S$ are expressed now as
$S=S(x_+,x_-)=\sum_{k,n}s_{kn}e^{ikx_+}\cos nx_-$ where $k,n$ are
integers of the same parity, and $n\geq0$.

{}From (\ref{eq:def-M}),
(\ref{eq:def-ker-M}) we obtain
\begin{eqnarray*}
\lefteqn{\tilde S(y_+,y_-)=
 \kappa\left(\sin^2\frac{y_+}{2}-\sin^2\frac{y_-}{2}\right)^{-3g+1}\times}\\
&&\times\int_{y_-}^{y_+}dx_-\,\sin x_-
 \left(\sin^2\frac{x_-}{2}-\sin^2\frac{y_-}{2}\right)^{g-1}
 \left(\sin^2\frac{y_+}{2}-\sin^2\frac{x_-}{2}\right)^{2g-1}
 S(y_+,x_-).
\end{eqnarray*}

Let us make now the change of variables
\beq
 \xi_\pm=\sin^2\frac{x_\pm}{2}, \quad
   d\xi_\pm=\half\sin x_\pm\,dx_\pm, \qquad
   \eta_\pm=\sin^2\frac{y_\pm}{2},
\eeq
denoting $\check S(x_+,\xi_-)=S(x_+,x_-)$. It is easy to see that
$\check S(x_+,\xi_-)$ is polynomial in $\xi_-$ and that
\beq\label{eq:SS}
 \tilde S(y_+,y_-)=
 2\kappa(\eta_+-\eta_-)^{-3g+1}\int_{\eta_-}^{\eta_+}d\xi_-\,
    (\xi_--\eta_-)^{g-1}(\eta_+-\xi_-)^{2g-1}\check S(y_+,\xi_-).
\eeq

Now put
$$ \xi_-=(\eta_+-\eta_-)\xi+\eta_- $$
and choose
\beq\label{def-kappa}
 \kappa=\frac{1}{2B(g,2g)}=\frac{\Gamma(3g)}{2\Gamma(g)\Gamma(2g)}.
\eeq

Then, finally
\beq\label{eq:S-final}
 \tilde S(y_+,y_-)=
   \frac{\Gamma(3g)}{\Gamma(g)\Gamma(2g)}
   \int_0^1 d\xi\, \xi^{g-1}(1-\xi)^{2g-1}
   \check S(y_+,(\eta_+-\eta_-)\xi+\eta_-).
\eeq

It is sufficient to calculate the integral (\ref{eq:S-final})
for the monomials
$$ \check S=e^{iky_+}\eta_-^l(\eta_+-\eta_-)^m\xi^m $$
such that $k,l,m\in\Z$, $l,m\geq 0$ and $k\equiv l+m \pmod{2}$.
Evaluating the beta-function integral
$$
  \int_0^1 d\xi\, \xi^{g-1+m}(1-\xi)^{2g-1}=
  \frac{\Gamma(g+m)\Gamma(2g)}{\Gamma(3g+m)}
$$
one obtains
\beq\label{eq:M-basis}
 \tilde S(y_+,y_-)=\frac{\Gamma(3g)\Gamma(g+m)}{\Gamma(3g+m)\Gamma(g)}
e^{iky_+}\eta_-^l(\eta_+-\eta_-)^m.
\eeq

It is easy to verify that the result is a symmetric trigonometric
polynomial in $y_1$, $y_2$. \endproof

Note that the normalization constant $\kappa$ is chosen in such a way
that $M:1\mapsto 1$.

The formula (\ref{eq:M-basis}) shows that the operator $M$
can in fact be continued analytically in $g$ on the whole complex
plane excluding the points $g=-\half,-1,-\frac{3}{2},\ldots$
coming from the poles of the gamma functions in (\ref{eq:M-basis})
and also
$g=-\frac{1}{3},-\frac{2}{3},\ldots$ coming from the poles
of $\Gamma(3g)$ in the normalization constant $\kappa$ (\ref{def-kappa}).

\section{$A_2$: Separated equation}\label{sep-eq}

To complete the proof of the theorem \ref{Psi3-fact}
we need to learn more about the separated equation (\ref{eq:sep-eq3}).

Eliminating from $\psi$ the vacuum factor $ \omega(y)=\sin^{2g}y $
via the substitution $\psi(y)=\phi(y)\omega(y)$ one obtains
\begin{eqnarray}\label{eq:phi-eq-3}
 \lefteqn{\left[i\dd_y^3+
(h_1+6ig\cot y)\dd_y^2\right.} \nonumber \\
&&+(-i(h_2+12g^2)+4gh_1\cot y+3ig(3g-1)\sin^{-2}y)\dd_y \nonumber \\
   &&+\left.(-(h_3+4g^2h_1)-2ig(h_2+4g^2)\cot y
      +g(3g-1)h_1\sin^{-2}y)\right]\phi=0.
\label{eq:eq-phi3}
\end{eqnarray}

The change of variable $t=e^{2iy}$ brings the last equation to
the Fuchsian form:
\beq\label{eq:Fuchs3}
  \left[\dd_t^3+w_1\dd_t^2+w_2\dd_t+w_3\right]\phi=0
\eeq
where
\begin{eqnarray*}
w_1&\!\!=\!\!&-\frac{3(g-1)+\half h_1}{t}
        +\frac{6g}{t-1}, \\
w_2&\!\!=\!\!&\frac{(3g^2-3g+1)+\half(2g-1)h_1+\frac{1}{4}h_2}{t^2}
 +\frac{3g(3g-1)}{(t-1)^2}
 -\frac{g(9(g-1)+2h_1)}{t(t-1)}, \\
w_3&\!\!=\!\!&-\frac{g^3+\half g^2h_1+\frac{1}{4}gh_2+\frac{1}{8}h_3}{t^3}
 +\frac{\half g((h_2+4g^2)(t-1)-(3g-1)h_1)}{t^2(t-1)^2}.
\end{eqnarray*}

The points $t=0,1,\infty$ are regular singularities with the exponents
$$ \begin{array}{lll}
 t\sim 1\quad & \phi\sim(t-1)^\mu\quad & \mu\in\{-3g+2,-3g+1,0\} \\
 t\sim 0\quad & \phi\sim t^\rho\quad & \rho\in\{n_1,n_2+g,n_3+2g\} \\
 t\sim\infty\quad & \phi\sim t^{-\s}\quad & -\s\in\{n_1-2g,n_2-g,n_3\}
\end{array}
$$

Like in the $A_2$ case, the equation (\ref{eq:Fuchs3}) is reduced
by the substitution $\phi(t)\!=\!t^{n_1}(1-t)^{1-3g}f(t)$
to the standard ${}_3F_2$ hypergeometric form \cite{Bateman}
\beq\label{eq:hge-32}
  \left[t\dd_t(t\dd_t+b_1-1)(t\dd_t+b_2-1)
 -t(t\dd_t+a_1)(t\dd_t+a_2)(t\dd_t+a_3)\right]f=0,
\eeq
the parameters $a_1$, $a_2$, $a_3$, $b_1$, $b_2$ being given by the formulas
(\ref{eq:def-hgf}) which for $N=3$ read
$$ a_1=n_1-n_3+1-3g, \qquad a_2=n_1-n_2+1-2g, \qquad
   a_3=1-g,
$$
$$ b_1=n_1-n_3+1-2g, \qquad b_2=n_1-n_2+1-g. $$

\begin{prop}\label{Laurent-3}
 Let the parameters $h_k$ be given by (\ref{eq:def-h3}),
(\ref{eq:def-m3}) for a
triplet of integers $\{n_1\leq n_2\leq n_3\}$ and $g\neq
1,0,-1,-2,\ldots$. Then the equation (\ref{eq:Fuchs3}) has
a unique, up to a constant factor, Laurent-polynomial solution
\beq
 \phi(t)=\sum_{k=n_1}^{n_3}t^k c_k(\vec n;g),
\eeq
the coefficients $c_k(\vec n;g)$ being rational functions of
$k$, $n_j$ and $g$.
\end{prop}

The above proposition follows from a more general statement.

\begin{theo}\label{Vadim}
Let the function $F_{n_1,\ldots,n_N}(t)$ be given for $|t|<1$ by
the right hand side of the formula (\ref{eq:phi-hgf}), the parameters
$a_j$ and $b_j$ being given by (\ref{eq:def-ab}) for some sequence
of integers $\vec n=\{n_1\leq n_2\leq\ldots \leq n_N\}$.
Let $g\neq 1,0,-1,-2,\ldots$. Then $F_{\vec n}(t)$ is a Laurent
polynomial
\beq
 F_{\vec n}(t)=\sum_{k=n_1}^{n_N}t^k c_k(\vec n;g),
\eeq
the coefficients $c_k(\vec n;g)$ being rational functions of
$k$, $n_j$ and $g$.
\end{theo}

{\bf Proof.} Consider first the hypergeometric series (\ref{eq:def-hgf})
for ${}_NF_{N-1}$ which converges for $|t|<1$.
Using for $a_j$ and $b_j$ the
expressions (\ref{eq:def-ab}) one notes that
$a_{j+1}=b_j+n_{N-j+1}-n_{N-j}$ and therefore
$$
  \frac{(a_{j+1})_k}{(b_j)_k}=
  \frac{(b_j+k)_{n_{N-j+1}-n_{N-j}}}{(b_j)_{n_{N-j+1}-n_{N-j}}}.
$$

The expression
$$
 \frac{(a_2)_k\ldots(a_N)_k}{(b_1)_k\ldots(b_{N-1})_k}=
 \frac{(b_1+k)_{n_N-n_{N-1}}\ldots(b_{N-1}+k)_{n_2-n_1}}%
{(b_1)_{n_N-n_{N-1}}\ldots(b_{N-1})_{n_2-n_1}}=
P_{n_N-n_1}(k)
$$
is thus a polynomial in $k$ of degree $n_N-n_1$.
So we have
$$
 {}_NF_{N-1}(a_1,\ldots,a_{N};b_1,\ldots,b_{N-1};t)=
\sum_{k=0}^\infty \frac{(a_1)_k}{k!}P_{n_N-n_1}(k)t^k
$$
from which it follows that
$$
  {}_NF_{N-1}(a_1,\ldots,a_{N};b_1,\ldots,b_{N-1};t)=
  \tilde P_{n_N-n_1}(t)(1-t)^{Ng-1}
$$
where $\tilde P_{n_N-n_1}(t)$ is a polynomial of degree $n_N-n_1$
in $t$. \endproof

To prove now the proposition \ref{Laurent-3} it is sufficient to
notice that in the case $N=3$
the hypergeometric series ${}_3F_2(a_1,a_2,a_3;b_1,b_2;t)$
satisfies the same equation (\ref{eq:hge-32}) as $f(t)$ and
therefore the Laurent polynomial $F_{\vec n}(t)$ constructed
above satisfies the equation (\ref{eq:Fuchs3}).
The uniqueness follows from the fact that all the linearly
independent solutions to (\ref{eq:Fuchs3}) are nonpolynomial which
is seen from the characteristic exponents.
\endproof

Now everything is ready to finish the proof of the theorem \ref{Psi3-fact}.
Since the function $\tilde J_{n_1n_2n_3}(y_1,y_2;Q)$ satisfies
(\ref{eq:phi-eq-3}) in variables $y_{1,2}$ and is a Laurent polynomial
it inevitably has the factorized form
\beq\label{eq:fact-J}
 \tilde J_{n_1n_2n_3}(y_1,y_2;Q)=
e^{ih_1Q}\phi_{n_1n_2n_3}(y_1)\phi_{n_1n_2n_3}(y_2)
\eeq
 by virtue of the proposition \ref{Laurent-3}. \endproof

\section{Integral representation for Jack polynomials}\label{inv-K}

The formula (\ref{eq:fact-J}) presents an interesting
opportunity to construct a new integral representation of the
Jack polynomial $J_{\vec n}$ in terms of the ${}_3F_2$
hypergeometric polynomials $\phi_{\vec n}(y)$ constructed above.
To achieve this goal, it is necessary to invert explicitely the
operator $M:J\mapsto\tilde J$.

Let us examine again the integral (\ref{eq:SS}).
Assume that $x_+=y_+$ and respectively $\xi_+=\eta_+$ are fixed
whereas $\xi_-$, $y_-$ are variables. Then, denoting
$$ \tilde s(\eta_-)=
\frac{1}{2\kappa\Gamma(g)}\tilde S(y_+,y_-)(\eta_+-\eta_-)^{3g-1},
\quad s(\xi_-)=(\eta_+-\xi_-)^{2g-1}\check S(y_+,\xi_-)
$$
we face the problem of inverting the integral transform
\beq\label{eq:frac-int}
     \tilde s(\eta_-)=\int_{\eta_-}^{\eta_+}d\xi_-\,
    \frac{(\xi_--\eta_-)^{g-1}}{\Gamma(g)}s(\xi_-)
\eeq
which is known as Riemann-Liouville integral of fractional order
$g$ \cite{Int-tr}. Its inversion is formally given  by changing sign
of $g$
\beq
      s(\xi_-)=\int_{\xi_-}^{\xi_+}d\eta_-\,
    \frac{(\eta_--\xi_-)^{-g-1}}{\Gamma(-g)}\tilde s(\eta_-)
\eeq
and is called fractional differentiation operator.
However, by our assumption $g>0$, so the integrand becomes
singular at $\xi_-=\eta_-$ and the integral should be
regularized in the standard way \cite{G-Sh}.

Retracing all the intermediate transformations we
obtain
$$ S(x_+,x_-)=
 \frac{\Gamma(2g)}{\Gamma(-g)\Gamma(3g)}
 (\xi_+-\xi_-)^{-2g+1}\int_{\xi_-}^{\xi_+}d\eta_-\,
(\eta_--\xi_-)^{-g-1}(\xi_+-\eta_-)^{3g-1}
\tilde S(x_+,y_-)
$$
and finally come to the formula for $M^{-1}:\tilde J\mapsto J$

\beq\label{eq:JMJ}
 J(x_+,x_-;Q)=\int_{x_-}^{x_+}dy_-\,
    \check{\cal M}(x_+,x_-;y_-)\tilde J(x_+,y_-;Q)
\eeq
\beq\label{eq:inv-M}
  \check{\cal M}=\check\kappa
\frac{\sin y_-
 \left[
   \sin\left(\frac{x_++y_-}{2}\right)
   \sin\left(\frac{x_+-y_-}{2}\right)
 \right]^{3g-1}}%
{\left[
   \sin\left(\frac{y_-+x_-}{2}\right)
   \sin\left(\frac{y_--x_-}{2}\right)
 \right]^{g+1}
\left[\sin x_1\sin x_2\right]^{2g-1}}
\eeq
where
\beq
 \check\kappa=\frac{\Gamma(2g)}{2\Gamma(-g)\Gamma(3g)}.
\eeq

For $K^{-1}$ we have respectively
\beq
  \check{\cal K}=\check\kappa
\frac{\sin^g x_-\sin y_-
 \left[
   \sin\left(\frac{x_++y_-}{2}\right)
   \sin\left(\frac{x_+-y_-}{2}\right)
 \right]^{g-1}}%
{\left[
   \sin\left(\frac{y_-+x_-}{2}\right)
   \sin\left(\frac{y_--x_-}{2}\right)
 \right]^{g+1}
\left[\sin x_1\sin x_2\right]^{g-1}}.
\eeq

The formulas (\ref{eq:fact-J}), (\ref{eq:JMJ}), (\ref{eq:inv-M})
provide a new integral representation for Jack polynomial
$J_{\vec n}$ in terms of the ${}_3F_2$
hypergeometric polynomials $\phi_{\vec n}(y)$. The representation
would acquire more satisfactory form if one could describe
explicitely the normalization of $\phi$ corresponding to the
standard normalization (\ref{eq:norm-J}) of $J$. We intend to study
this question in a subsequent paper.

It is remarkable that for positive integer $g$ the operators
$K^{-1}$, $M^{-1}$ become differential operators
of order $g$. In particular, for $g=1$ we have $K^{-1}=\dd/\dd y_-$.

\section{Separation of variables in the Schur polynomials}
\label{Schur}

For the generic $g$ the separation of variables
in Jack polynomials is so far unknown for $N>3$. However,
the problem simplifies drastically in the
case $g=1$, when Jack polynomials are reduced to the Schur polynomials
\cite{Macd}, and allows quite simple solution. In the present
section we have changed notation to make it more convenient
for handling Schur polynomials.

Let
\beq
 P_{n_1\ldots n_N}(t_1,\ldots,t_N)=\det
   \left|\begin{array}{cccc}
      t_1^{n_1} & t_2^{n_1} & \ldots & t_N^{n_1} \\
      t_1^{n_2} & t_2^{n_2} & \ldots & t_N^{n_2} \\
      \ldots & \ldots & \ldots & \ldots \\
      t_1^{n_N} & t_2^{n_N} & \ldots & t_N^{n_N}
    \end{array}\right|.
\eeq

Schur polynomial is defined as the ratio of two antisymmetric
polynomials:
\beq
  S_{\vec n}(\vec t)=\frac{P_{n_1,n_2+1,\ldots,n_N+N-1}(\vec t)}%
{P_{0,1,2,\ldots,N-1}(\vec t)}.
\eeq

Denominator (corresponding to $\Omega$ in the previous sections)
\beq
  P_{0,1,2,\ldots,N-1}(\vec t)=\prod_{k>j}(t_k-t_j)
\eeq
is the elementary antisymmetric polynomial (Vandermonde determinant).

The separated equation
\beq
 \prod_{j=1}^N \left(t\dd_t-n_j\right)\psi(t)=0.
\eeq
has as the general solution the polynomial
$ \psi(t)=\sum_{j=1}^N c_j t^{n_j}$.
The boundary condition
\beq
 \left.\frac{\dd^k}{\dd t^k}\psi(t)\right|_{t=1}=0, \qquad
 k=0,1,\ldots,N-2
\eeq
selects the solution
\beq
 c_j\sim \det\left|\begin{array}{cccccc}
     1 & \ldots & 1 & 1 & \ldots & 1 \\
     n_1 & \ldots & n_{j-1} & n_{j+1} & \ldots & n_N \\
     n_1^2 & \ldots & n_{j-1}^2 & n_{j+1}^2 & \ldots & n_N^2 \\
   \ldots & \ldots & \ldots & \ldots & \ldots & \ldots \\
     n_1^{N-2} & \ldots & n_{j-1}^{N-2} & n_{j+1}^{N-2} & \ldots & n_N^{N-2}
        \end{array}\right|
  =\prod_{\scriptstyle k>l \atop \scriptstyle k,l\neq j}(n_k-n_l).
\eeq

In case of Schur polynomials it is easier to construct
the inverse operator $K^{-1}$ rather than $K$. Let
\beq
 \tilde\Psi(t_1,\ldots,t_{N-1})=\psi(t_1)\ldots\psi(t_{N-1})
      =\prod_{j=1}^{N-1}\psi(t_j)
\eeq
and
\beq
 K^{-1}=\prod_{k>j}\left(t_k\dd_{t_k}-t_j\dd_{t_j}\right).
\eeq

\begin{theo}
The operator $K^{-1}$ transforms the symmetric polynomial
$\tilde\Psi$ into an antisymmetric polynomial
$\Psi(t_1,\ldots,t_{N-1})=K^{-1}\tilde\Psi$
which is none other than the numerator of Schur polynomial
\beq
 \Psi\left(\frac{t_1}{t_N},\ldots,\frac{t_{N-1}}{t_N}\right)
 t^{n_1+\cdots+n_N}
\sim P_{n_1\ldots n_N}(t_1,\ldots,t_N).
\eeq
\end{theo}

The proof consists in an elementary calculation.

Since we have already seen in the $N=3$ case that $K^{-1}$ becomes
a differential operator for integer $g>0$,
it is not surprising that here $K^{-1}$ is also a differential operator.

\section{Discussion}

The construction of the operator $M$ performing the separation of
variables for Jack polynomials originates from mathematical physics
(Inverse Scattering Method) and contains a lot of guesswork.
A generalization of our results to the case of higher rank
$N>3$ could probably throw some light
on the algebraic and geometric meaning of the whole
construction which remains still obscure. The only available
results in this direction are so far the case $g=1$ (Schur
polynomials) and theorem \ref{Vadim} which allows to formulate
conjecture 2 about the structure of separated polynomials in the
general case.

Among other challenging problems one should mention generalizations
to other root systems, first of all $BC_N$, and also to the
$q$-finite-difference case (Macdonald polynomials).

\bigskip
\noindent{\it Acknowledgments.}
VK acknowledges support by the Nederlandse Organisatie voor
Wetenschappelijk Onderzoek (NWO).
ES wishes to thank
K.~Aomoto and S.G.~Gin\-di\-kin
for their interest in the work and valuable remarks.

\end{document}